\documentclass[aps,twocolumn,pra,showpacs,floatfix]{revtex4-2}
\usepackage{epsfig}
\usepackage{graphicx}
\usepackage{dcolumn}
\usepackage{amssymb,amsmath}
\usepackage{mathrsfs}
\begin{document}

\title{Using the Th~III Ion for a Nuclear Clock and Searches for New Physics}

\author{V. A. Dzuba and V. V. Flambaum}

\affiliation{School of Physics, University of New South Wales, Sydney 2052, Australia}

\begin{abstract}

The $^{229}$Th nucleus possesses a unique low-frequency  transition at 8.4 eV, which is being considered for the development of an extremely accurate nuclear clock. We investigate an electronic bridge process  in  the Th~III ion,  where nuclear excitation is mediated  by electronic transitions,  and show that selecting appropriate laser frequencies can enhance the nuclear excitation probability by up to $10^4$  times.  Electrons also reduce 1.7 times the lifetime of the nuclear excited state.

Moreover, the electronic structure of Th III  offers exceptional opportunities for probing new physics. Notably, the ion contains a metastable electronic state coupled to the ground state via a weak M2 transition, which can be utilised for quantum information processing, as well as searches for oscillating dark matter axion and scalar fields, violation of local Lorentz invariance, test of the Einstein's equivalence principle, and measurement of the nuclear weak quadrupole moment.  Additionally, the electronic states of Th III  present a unique case of level crossing involving the $5f$, $6d$, and $7s$ single-electron states. This crossing renders high sensitivity of the transition frequencies to potential time-variation of the fine-structure constant.

  \end{abstract}

\maketitle

\section{Introduction}

The nucleus of the $^{229}$Th isotope has a unique feature - a low-lying excited state~\cite{Th1} connected to the ground state via a nuclear M1 transition. It has been proposed to use this transition as the basis for a nuclear clock~\cite{Th2} with exceptionally high accuracy (up to $\sim 10^{-19}$~\cite{Kuzmich}), which is also highly sensitive to new physics~\cite{Flambaum,Flambaum2016}. 

The small size of the nucleus and the shielding effect of atomic electrons render the transition frequency  largely insensitive to perturbations when the Th ion is in a "stretched" state, i.e., with maximal projections of both electronic and nuclear angular momenta~\cite{Kuzmich}.  
 On the other hand, the small nuclear transition energy (8.4 eV) arises from a strong cancellation between contributions from strong and electromagnetic nuclear forces. This makes the transition frequency highly sensitive to small changes in each of these contributions, which may result from a hypothetical variation of fundamental physical constants, such as the fine-structure constant $\alpha$, strong interaction constants, and quark masses~\cite{Flambaum}. 

Furthermore, this transition is highly sensitive to dark matter fields, which could induce variations in these constants (see, e. g., ~\cite{Arvanitaki,Stadnik}). For example, the effects of the interaction between scalar field dark matter and fermions may be presented as the apparent variation of fermion masses. This  follows from a comparison of the interaction of a fermion with the scalar field $ -g_f m_f  \phi^n \bar \psi \psi $ and the fermion mass term in the Lagrangian $- m_f \bar \psi \psi$. Adding these terms gives $m'_f=m_f(1+ g_f  \phi^n)$, where $n=1,2$. Similarly, the interaction of scalar dark matter with the electromagnetic field may be presented  as a variable fine structure constant $\alpha'=\alpha (1 + g_{\gamma} \phi^n)$.  

Note that if the interaction is quadratic in $\phi$, we may replace the scalar field by the pseudo-scalar (axion) field as $\phi^2$ always has positive parity \cite{Stadnik}. The corresponding theory  for QCD axion has been developed in Ref. \cite{Perez2022}%, in which limits on the axion interaction from atomic spectroscopy experiments were obtained 
(see also \cite{Samsonov}).

 The $^{229}$Th nuclear transition is also highly sensitive to potential violations of the Lorentz invariance and the Einstein's equivalence principle~\cite{Flambaum2016}.

The measurement of the energy of the nuclear clock transition in $^{229}$Th has been an ongoing effort for many years~\cite{wn1,wn2,wn3}, with significant recent progress~\cite{Th-wn0,Th-wn,Th-wn1}. The transition energy has been measured to be $\omega_N = 8.355733(2)_{\rm stat}(10)_{\rm sys}$~eV (67393 cm$^{-1}$)~\cite{Th-wn} and $\omega_N = 2,020,407,384,335(2)$~kHz~\cite{Th-wn1} in Th atoms inside solids.  A higher accuracy is expected in ion clocks ~\cite{Kuzmich}.  

The amplitude of the nuclear M1 transition is suppressed by five orders of magnitude compared to typical atomic E1 transitions. It is possible  to enhance this transition using  the electronic bridge (EB) process, in which electronic transitions induce nuclear transitions via the hyperfine interaction  ~\cite{Tkalya1,Tkalya2}. The EB effect also increases the probability of decay of the nuclear excited state~\cite{Strizhov}. Laser-induced internal conversion in $^{229}$Th ion was suggested in Ref. \cite{Karpeshin1992}. Different versions of EB processes  for  the  isomeric nuclear  state decay and excitation have been explored in several Th II, Th III, and Th IV  calculations, see e.g. Refs.  ~\cite{EB1,EB2,EB3,EB4,EB5,EB6,EB7,EB8,EB9,EB11,EB12}.
%In particular, Ref.~\cite{EB3} proposed using the EB process for the resonance nuclear excitation in Th~II. 

In this work, we investigate  EB process for the Th~III ion, which may offer several advantages over Th~IV and Th~II. The electronic spectrum of Th~III is  sufficiently rich  to enable strong enhancement of the EB effect through a two-step process, similar to that suggested in Ref.~\cite{EB3} for Th~II. At the same time, its relatively simple electronic structure, consisting of two valence electrons above a closed-shell core, allows for more accurate calculations and a clearer interpretation of experimental results. {We also considered an example of a single-step process where the Th III ion is excited by a single high energy  photon directly from the ground state.}

In contrast, the electronic bridge mechanism does not produce any enhancement in Th~IV. In Th~II the spectrum is very dense, making it  impossible to  reliably match experimental energy levels with the calculated energy levels and  their corresponding wave functions.  While the enhancement in Th~II is expected to be strong, its precise magnitude cannot be reliably estimated.  

Another advantage of using the Th~III ion is the presence of  an  electronic clock transition.  The first excited state of Th~III is a low-lying metastable state (excitation energy $\approx 63$ cm$^{-1}$) connected to the ground state via a weak M2 transition.   Due  to the extremely long lifetime of the upper state, it effectively serves as a second ground state of the ion. 

Notably, searches for new physics using atomic clocks typically rely on comparing the frequencies of two different clock transitions. Having both clock transitions, nuclear and atomic,  within the same system provides a significant advantage for such measurements.
 
 Additionally, this doublet of states can be utilized for quantum information applications, similar to proposal involving a highly charged ion, as suggested in Ref.~\cite{Safronova}. Th III  offers a more practical alternative, as it is significantly easier to produce and potentially simpler  to operate.  Here  we specifically refer to the natural   isotope $^{232}$Th.
 
% Previously, a system with two ground states was proposed for quantum information applications in a highly charged ion (see Ref. [20]). Th III  offers a more practical alternative, as it is significantly  easier to produce and potentially simpler to operate.  Here we specifically refer to the  stable  isotope $^{232}$Th.
 
 The electronic clock transition can be used to   search for the oscillating axion field. The axion field may interact directly with electrons or induce an oscillating nuclear magnetic quadrupole moment, which, in turn, stimulates the M2 transition. Unlike photon-induced  M2 transitions, axion-induced M2 transitions are not suppressed, resulting in a significant relative reduction in background noise~\cite{axion}.

The ground state of Th~III is also promising for testing the local Lorentz invariance (LLI) violation. Its large total electron angular momentum and the presence of an electron in the open $5f$ subshell ensure an enhancement  of the  LLI violation effect - see detailed explanation  in Ref. ~\cite{Nature}. The M2 transition can further be used to search for violations of the Einstein's equivalence principle (EEP), which may manifest via an annual modulation of atomic frequencies due to the varying  distance to Sun and corresponding variation  of its gravitational potential  - see e.g. ~\cite{Kostelecky1,Kostelecky2,Kostelecky3,Shuryak,EEP}. 

Additionally, Th~III exhibits a unique case of multiple level crossing. The energies of the $5f$, $6d$, and $7s$ single-electron states are approximately equal, making the transition frequencies particularly sensitive to potential time variation of the fine-structure constant $\alpha$ - see explanation in Refs.  ~\cite{Porsev2009,HCI,crossing}. This sensitivity is especially pronounced for the aforementioned M2 transition, as the small value of its frequency $\omega$ leads to an  enhancement of the relative effect  $\delta \omega /\omega$.

Finally, the ground state of Th~III (with total electron angular momentum $J=4$ and negative parity) is mixed with the metastable $J=2$ positive-parity state at 63 cm$^{-1}$ via the interaction of electrons with the nuclear weak quadrupole moment. The small 63 cm$^{-1}$ energy denominator leads to an enhancement of corresponding parity-violating effects. Measuring these effects would enable, for the first time, the determination of the quadrupole moment of the neutron distribution in nuclei, which provides the dominant contribution to the weak quadrupole moment~\cite{Sushkov,Khriplovich,Harabati,Lackenby}.

%In this paper, we present accurate calculations of the discussed effects in the Th~III ion. Our analysis includes the EB processes in the excitation and decay of the 8.4 eV nuclear clock state, the sensitivity of atomic frequencies to variations in the fine-structure constant $\alpha$, and other effects relevant to nuclear clock development and the search for new physics.

\begin{table} 
  \caption{\label{t:ener}States of interest for  the electronic bridge process in Th~III. Abbreviation GS stands for the ground state.
  Other states are named in accordance with diagram of Fig.~\ref{f:eb}, i.e., states T1, T2 and T3 appear on the line $t$, states N1, N2, N3, N4, N5 - on the line $n$, states S1, S2, S3, S4, S5 - on the line $s$. }
\begin{ruledtabular}
\begin{tabular}   {lll c rl rl}
\multicolumn{1}{l}{State}&
\multicolumn{1}{l}{Confi-}&
\multicolumn{1}{l}{Term}&
\multicolumn{1}{c}{$J$}&
\multicolumn{2}{c}{NIST~\cite{NIST}}&
\multicolumn{2}{c}{This work}\\
\multicolumn{1}{c}{name}&
\multicolumn{1}{c}{guration}&
%\multicolumn{1}{c}{Term}&
%\multicolumn{1}{c}{$J$}&
&&\multicolumn{1}{c}{Energy}&
\multicolumn{1}{c}{Land\'{e}}&
\multicolumn{1}{c}{Energy}&
\multicolumn{1}{c}{Land\'{e}}\\
&&&&\multicolumn{1}{c}{[cm$^{-1}$]}&&
\multicolumn{1}{c}{[cm$^{-1}$}]& \\
\hline
GS & $5f6d$ & $^3$H$^{\rm o}$ & 4 & 0 & 0.888  & 0 & 0.885 \\ 
T1 & $5f^2$ & $^1$G         & 4 & 25972 & 1.072 & 29323 & 1.11 \\ %? e4 4
T2 & $5f7p$ & $(7/2,1/2)$  & 4 & 38580 & 1.105 & 38190 & 1.107 \\
T3 & $5f7p$ & $(5/2,3/2)$  & 4 &  43702 & 1.069 & 44354   & 1.072   \\
T4 & $5f7p$ & $(5/2,3/2)$  & 3 &  42313 & 0.971 & 43184   & 0.968   \\

N1 & $5f8s$ & $(5/2,1/2)^{\rm o}$& 3 & 74784 &     & 74030 & 1.05 \\
N2 & $5f7d$ & $(5/2,3/2)^{\rm o}$& 3 & 78328 &     & 78263 & 0.820 \\
N3 & $5f8s$ &                  & 4 & 78417 &     & 78573 & 1.22 \\ %? o4 8 or 7 or 9
N4 & $5f7d$  & $(7/2,3/2)^{\rm o}$ & 3 & 82827 & & 84812 & 1.146  \\
N5 & $5f8p$  & $(5/2,1/2)$ & 3 & 86086 & & 88337 & 0.855 \\

S1 & $5f8s$ &                            & 3 &  7501 & 1.027 &  6901 & 1.029 \\ % o3 3
S2 & $5f6d$ & $^3$D$^{\rm o}$ & 3 & 10741 & 1.22  & 13685 & 1.18 \\ % o3 5
S3 & $5f6d$ & $^1$F$^{\rm o}$ & 3 & 15453 & 1.07  & 18110 & 1.12 \\ % o3 6
S4 & $5f6d$ & $^3$F$^{\rm o}$ & 2 &  511 & 0.739 &   957 &  0.785 \\
S5 & $5f^2$ & $^3$F & 2 &  18864 & 0.694 &   20528 &  0.765 \\

\end{tabular}			
\end{ruledtabular}
%\footnotetext[1]{Ref. \cite{Th-mu-Q}.}
%\footnotetext[2]{Ref. \cite{ThQold}.}
%\footnotetext[3]{Ref. \cite{Thm}.}
\end{table}

\begin{table} 
  \caption{\label{t:S}Lifetimes ($\tau$) and width ($\Gamma=1/\tau$) of final states for the electronic bridge process in Th~III.}
\begin{ruledtabular}
\begin{tabular}   {lll c rcc}
\multicolumn{3}{c}{State}&
\multicolumn{1}{c}{$J$}&
\multicolumn{1}{c}{$E$ (cm$^{-1}$)}&
\multicolumn{1}{c}{$\tau$}&
\multicolumn{1}{c}{$\Gamma$}\\
\hline
S1 & $5f8s$ &                            & 3 &    7501 & 332 $\mu$s & 3.0 kHz \\
S2 & $5f6d$ & $^3$D$^{\rm o}$ & 3 & 10741 & 9.2 $\mu$s & 109 kHz \\
S3 & $5f6d$ & $^1$F$^{\rm o}$ & 3 & 15453 & 2.5 $\mu$s & 400 kHz \\
S4 & $5f6d$ & $^3$F$^{\rm o}$ & 2 &     511 & 57 ms & 17 Hz \\
S5 & $5f^2$ & $^3$F & 2 &   18864 & 0.10 $\mu$s & 10 MHz \\

\end{tabular}			
\end{ruledtabular}
\end{table}

%\begin{table} 
% \caption{\label{t:E1} Electric dipole (E1) amplitudes (reduced matrix elements $|\langle A||\mathbf{D}||B\rangle|$, a.u.) between states of interest presented in Table~\ref{t:ener}.}
%\begin{ruledtabular}
%\begin{tabular}{l ccc cccc}
% A \textbackslash  \ B  &   GS    &  N1     &    N2   &  N3  \\
 %\hline
%T1 &  1.1628   &  0.2420 &  0.2245 &   0.4254 \\
%T2 &   3.520   & 0.9187 &  1.0049 &   3.0468 \\
%\end{tabular}			
%\end{ruledtabular}
%\end{table}

\section{Electronic bridge for the  nuclear excitation  in Th~III}

In this section, we analyze the EB process for Th~III, focusing on resonance enhancement. 
Our approach follows the method used in Ref.~\cite{EB3} for the Th~II ion.  The decay of any atomic state with energy $E_n > \omega_N$ may include nuclear excitation. Since the nuclear excitation energy $\omega_N = 67393$ cm$^{-1}$ lies outside the optical region, we consider a two-step excitation of the electronic states. In the first step, the atom is excited from the ground state (GS) to an intermediate state $t$ with energy $\omega_1 \sim \omega_N / 2$.  We have found  three suitable states with electron angular momentum $J=4$, referred to as T1, T2 and T3 (see Table~\ref{t:ener}), which connected to the GS by a strong electric dipole (E1) transition. We have not found suitable  states  with $J=5$.  We exclude states with $J=3$ to avoid leakage into the metastable state $5f6d \   ^3$F$^{\rm o}_2$ (E = 511~cm$^{-1}$). 
 
\begin{figure}[tb]
	\epsfig{figure=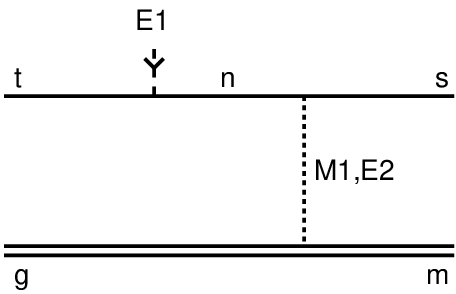,scale=0.6}
	\caption{Diagram for nuclear excitation  via  the electronic bridge  process in Th~III. The atom is initially in the { excited} electronic state $t$, which is populated by the first laser with frequency $\omega_1 = \epsilon_t$, {where $\epsilon_t$ is energy of electron excited state. The second  laser frequency is tuned to excite nucleus into the isomeric state. }  The E1 operator represents the interaction with the second laser, which has a frequency $\omega_2 = \omega_N + \omega_s - \omega_1$. Here, $n$ and $s$ denote the intermediate and final electronic states, respectively, while $g$ and $m$ correspond to the ground and isomeric nuclear states. The dotted line indicates the hyperfine interaction between atomic electrons and the nucleus.}
	\label{f:eb}
\end{figure}

In the second step, the ion is further excited by a second laser to satisfy the energy conservation condition for simultaneous nuclear excitation and excitation of the final electronic state, $\omega_1 + \omega_2 = \omega_N + \omega_s$. Note that  the final state $s$ is not the ground state but a low-lying excited state, as this configuration  provides the largest EB amplitude.  In this process, the intermediate electronic state with energy $E_n$ is off-resonance, meaning it is virtually excited and subsequently decays, inducing nuclear excitation via hyperfine interaction (through magnetic dipole and electric quadrupole interactions). The diagram illustrating this process is shown in Fig.~\ref{f:eb} (see also Ref.~\cite{EB3}).

{  When calculating the EB process,  we perform  summation over all intermediate states $n$. %including states far above ionization potential. 
However, our calculations show that the summation usually  is strongly dominated by few terms (in most cases one or two), which have small energy denominator and large matrix elements.
 Possible choices for the states $t$,  $s$ and intermediate states $n$, which correspond to the smallest energy denominator   in Eq. (\ref{e:g2a}), are presented in Table~\ref{t:ener}.}

The rate of an induced excitation  from state $b$ to state $a$, $W_{ba}^{\rm in}$,  can be calculated using the rate of a spontaneous transition $a \rightarrow b$,  $W_{ab}$  \cite{Sobelman}:
\begin{equation}\label{e:ab-ba}
 W_{ba}^{\rm in} = W_{ab}\frac{4\pi^3c^2}{\omega^3} I_{\omega}.
 \end{equation}
 Here $I_{\omega}$ is the intensity of isotropic and unpolarized incident radiation. Index $b$ corresponds to the nuclear ground state and electron excited state, index $a$ corresponds to nuclear excited state and electronic ground or low energy excited state.
 The rate of a spontaneous transition via electronic bridge process is given by \cite{EB1}
 \begin{equation}\label{e:geb}
 W_{ab} = \frac{4}{9}\left(\frac{\omega}{c}\right)^3 \frac{|\langle I_g||M_k||I_m \rangle|^2}{(2I_m+1)(2J_t+1)}G_2^{(k)},
 \end{equation}
Keeping in mind the relation (\ref{e:ab-ba}), we assume that $\omega$ in (\ref{e:geb}) is the frequency of second excitation ($\omega \equiv \omega_2$) which is chosen to get into a resonance situation, $\omega_2=\epsilon_s + \omega_N -\omega_1$.
Factor $G_2$ in (\ref{e:geb}) depends on electrons only. It corresponds to upper line of Fig.~\ref{f:eb}. In principle, it has summation over complete set of intermediate states  $n$, see e.g. \cite{EB1}). However, assuming resonance situation and keeping only one strongly dominating term, we have
\begin{equation}\label{e:g2a}
G_2^{(k)} \approx \frac{1}{2J_n+1}\left[\frac{\langle s || T_k || n\rangle \langle n||D|| t \rangle}{\omega_{ns} - \omega_N}\right]^2.
\end{equation}
Here $T_k$ is the electron part of the hyperfine interaction operator (magnetic dipole (M1) for $k=1$ and electric quadrupole (E2) for $k=2$),
$D$ is the electric dipole operator (E1). States $n$ and $s$ are chosen to get close to a resonance, $\epsilon_n-\epsilon_s \equiv \omega_{ns} \approx \omega_N$.

It is convenient to present the results in terms of dimensionless ratios ($\beta$) of electronic transition rates to nuclear transition rates.
\begin{equation}\label{e:beta}
\beta_{M1}=\Gamma^{(1)}_{\rm EB}/\Gamma_N(M1), \ \  \beta_{E2}=\Gamma^{(2)}_{\rm EB}/\Gamma_N(E2).
\end{equation}
Here $\Gamma^{(k)}$ is given by (\ref{e:geb}) ($\Gamma^{(k)} \equiv W_{ab}$). Both parameters, $\beta_{M1}$ and $\beta_{E2}$ can be expressed via $G2$, (Eq.~(\ref{e:g2a}))~\cite{EB1,EB4}
\begin{eqnarray}\label{e:beta1}
&&\beta_{M1}= \left(\frac{\omega}{\omega_N}\right)^3\frac{G_2^{(1)}}{3(2J_s+1)}, \\ 
\label{e:beta2}
&&\beta_{E2}= \left(\frac{\omega}{\omega_N}\right)^3\frac{4G_2^{(2)}}{k_n^2(2J_s+1)}.
\end{eqnarray}
The $M1$ and $E2$ contributions can be combined into one effective parameter $\tilde \beta$ using the known ratio of the widths of the nuclear $M1$ and $E2$ transitions, $\gamma = \Gamma_{\gamma}(E2)/\Gamma_{\gamma}(M1) = 6.9 \times 10^{-10}$, \cite{EB4}.
%PRC 97, 044320
Then $\tilde \beta = \beta_{M1}(1+\rho)$, where $\rho = \gamma \beta_{E2}/\beta_{M1}$.

The results of the calculations for a number of possible transitions are presented in Table~\ref{t:betan}. We see that the probability of the nuclear excitation may be enhanced up to $\mathbf{10^4}$ times. This enhancement may be achieved by proper choices of laser frequencies in a two-step process of atomic excitation which is followed by the nuclear excitation. 
In {Table~\ref{t:betan} we also present variants  of the EB process in which the final state is the ground state and has zero natural width. Here $\tilde \beta$ is small due to the absence of resonance terms in the summation over $n$ and small matrix elements of the hyperfine interaction.  We also considered a number of other transitions, e.g. all T1 - Si transitions (i=1,2,3).
In all these transitions  $\tilde \beta < 1$ and for this reason the results are not included in the table. 
The last line of the table corresponds to a single-photon nuclear excitation via EB process at large photon frequency.
This is the only single-photon transition we found with $\tilde \beta > 1$.}

\begin{figure}[tb]
	\epsfig{figure=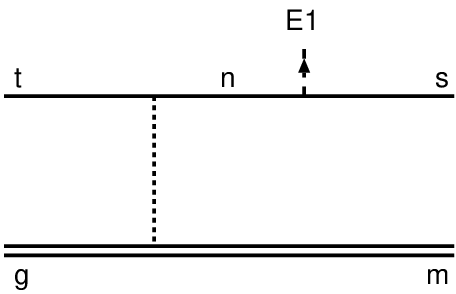,scale=0.6}
	\caption{Diagram for the  nuclear excitation via the decay of a highly excited atomic state $t$ with photon emission (E1).  {In this case the second laser excites atomic state with the energy $\epsilon_t$ exceeding the nuclear isomeric state energy, $\epsilon_t > \omega_N$. Then the excess  energy is taken away by the emitted photon.}  
Nuclear excitation is facilitated by the hyperfine interaction (dotted line).}  
\label{f:d}
\end{figure}

There is another possibility of the nuclear excitation via EB process suggested  in Ref.  \cite{EB7}, see Fig.~\ref{f:d}. If an atom is in excited state $t$ with energy larger that the nuclear excitation energy, $\epsilon_t > \omega_N$, then the decay of this state may include a channel with the  nuclear excitation via the hyperfine interaction while the excess of the energy is taken away by emitted photon. This contribution is strongly suppressed due to  a small value of the photon frequency,  $\omega=\epsilon_t-\epsilon_s-\omega_N$, since $\beta \sim \omega^3$, see Eqs. (\ref{e:beta1},\ref{e:beta2}).
For example, if $t$ is the state with $\epsilon_t$=83701~cm$^{-1}$ and $J=3$, then $\tilde \beta \approx 4\times 10^{-3}$.

\begin{table} 
\caption{\label{t:betan} Variants of the EB process with resonance nuclear excitation. 
Notations $t$, $n$, $s$ refer to the diagram on Fig.~\ref{f:eb}. State names Ti, Ni, Si (i=1,2,3,4,5) correspond to the states in Table~\ref{t:ener}.
We presented here only states Ni which  correspond to terms in the summation over $n$ with the minimal energy denominator ($\Delta_n$).
$\omega_1$ and $\omega_2$ are the frequencies of the first and second excitations respectively.
$\omega_1=\epsilon_f$, $\omega_2=\omega_N+\epsilon_s-\epsilon_f$;
$\Delta_n=\epsilon_n-\epsilon_s-\omega_N$.
Final state in the variants  8-10 is the ground state. Here $\tilde \beta$ is small due to  absence of the resonance terms and small  matrix elements of the hyperfine interaction.
%(numbers in square brackets stand for powers of ten).
The last line presents data for a single-photon nuclear excitation. }

%$\Delta$ is the energy denominator of the largest contribution. } 
\begin{ruledtabular}
\begin{tabular}{cccccc rr}

& $t$ &$n$ &$s$ & $\omega_1$ & $\omega_2$ & $\Delta_n$ & $\tilde \beta$ \\
&        &       &      &  [cm$^{-1}$] & [cm$^{-1}$] & [cm$^{-1}$] & \\
\hline

1 & T2 & N1 & S1 & 38580 & 36314 &  -110 &    186 \\
2 & T2 & N2 & S2 & 38580 & 39554 &   193 &    147 \\
3 & T2 & N4 & S3 & 38580 & 44266 &   -19 &  10014 \\
4 & T3 & N1 & S1 & 43702 & 31192 &  -110 &    557 \\
5 & T3 & N2 & S2 & 43702 & 34432 &   193 &     32 \\
6 & T3 & N4 & S3 & 43702 & 39144 &   -19 &    859 \\
7 & T3 & N1 & S4 & 43702 & 24202 &  4864 &    4.7 \\

8 & T1 &    & GS & 25972 & 41621 &   &     0.0013\\
9 & T2 &    & GS & 38580 & 28813 &   &     0.017\\
10 & T3 &    & GS & 43702 & 23691 &   &    0.24 \\
11 & T4 &    & GS & 42313 & 25080 &   &    0.27 \\

12 & GS &  N5  & S5 & 86086 &  & -170  &   12 \\

\end{tabular}			
\end{ruledtabular}
\end{table}

It was noted in Ref. \cite{EB11} that a very large value of EB enhancement  factor $\beta$  does not always  lead to the  $\beta$ times  enhancement of the excitation rate. Indeed,  the photon capture resonance  cross section contains the total width  in denominator,  which may be dominated by the radiative width $\Gamma_{\gamma}$,  if it is very strongly enhanced by the EB process ($\Gamma_{\gamma}=(1+\beta) \Gamma_N$, where $\Gamma_N$ is the the decay width of the nuclear isomeric state in the bare nucleus). The resonance photon capture cross section is 
\begin{equation} \label{crosssection}  
\sigma \sim \lambda^2 \frac{\Gamma_{\gamma}\Gamma }{(\omega-\omega_0)^2 +\Gamma^2/4},
 \end{equation}
where $\lambda$ is the photon wavelength,  $\Gamma= \Gamma_{\gamma} + \Gamma_{Ni} + \Gamma_s+ \Delta_l$, $\Gamma_{Ni}$ is the width of the nuclear transition (including the EB enhancement), $\Gamma_s$ is the width of the final electron state and $\Delta_l$ is the width of the energy distribution of the  photons in the pumping  laser beam (or width of the frequency comb teeth). There may be other line broadening mechanisms. 

In the resonance $\sigma \propto \beta \Gamma_N/( \beta \Gamma_N  +  \Gamma_{Ni}+ \Gamma_s+ \Delta_l)$.  We see that in the limit $\beta \to \infty$ the enhancement in the center of the resonance disappears.

However, in Th III  there is $\beta$ times enhancement of the cross section. 
Indeed, the decay widths of the final electron states are  relatively  large since they are determined by E1 transitions - see Table \ref {t:S}.   
The decay width in the bare nucleus is   $\Gamma_N \approx 5 \times 10^{-4}$ Hz   \cite{Katori,Th-wn0,Th-wn,Th-wn1,Hiraki,Zhang,Kraemer,Tkalya,Morgan,Perera}  , therefore,  the width with EB enhancement 
$\Gamma_{Ni} \approx \beta \Gamma_N  \leq $5 Hz $ \ll \Gamma_s$, and only $\beta$ in the numerator in the cross section Eq. (\ref{crosssection})  is important. 
%For the  final electron state  $S1$  $5f8s$,  $J=3$  with energy   7501  cm$^{-1}$  the radiative width  $\Gamma_s= $ Hz.

% and $^5$F$_{5/2}$ with the energy  1521.90  cm$^{-1}$ , the radiative width  $\Gamma_s$   is very small, 0.1 Hz, since it is produced by  low frequency M1 and E2 transitions. For other final states in the Table \ref{t:beta}, E1 transitions are possible, and the width is from 10 to 1000 Hz. Note that all lifetimes are strongly affected by the configuration mixing. 
%$\Gamma_N \approx 5 \times 10^{-4}$ Hz   \cite{Katori})

 Moreover,  
 %during the frequency scanning procedure  we are not exactly in the resonance,  i.e.  $(\omega-\omega_0)^2 > \Gamma^2/4$.  The width of the  frequency comb teeth $\Delta_l$ may significantly exceed the natural width $\Gamma$. Finally, 
 the scanning time is reduced inversely proportional to the width, so effectively one should  consider the product of the cross section times the total width, so $\beta$ in the denominator cancels out.  In this case the EB enhancement factor $\beta$ is important for any large $\beta$.

Assuming that the effective resonance width for the excitation to the isomeric state is dominated by the spectral width of the laser radiation  $\Delta_l$, Ref. \cite{EB3}  estimated excitation rate in Th II as  $W_{ba}^{\rm in}$=10 s$^{-1}$ for $\beta=30$.  Looking to values of $\beta$ in  Table~\ref{t:betan}, we may conclude that for the same  parameters of the laser beam  the excitation rate  in Th III may be few orders of magnitude higher. The total scanning time in Ref. \cite{EB3} was estimated as $10^4$ s for the scanning interval 1 eV.  Now the frequency of the nuclear transition is known with a much higher accuracy, the scanning interval $\sim 10^{-6}$ eV, so  the scanning time is expected to be much smaller.

\section{Electronic bridge for the nuclear decay in Th~III}

The electronic bridge also decreases  the lifetime of the nuclear excited state. There is no single dominant contribution in this case. The summation over all intermediate electron states (including continuum)  and final electrons states gives    $\tilde \beta \approx 0.7$.

The half-life in  Th IV was measured to be $T_{1/2}=1400^{+600}_{-300}$ s \cite{Katori}.
% corresponding to the lifetime $\tau=2020^{+866}_{-433}$. 
This half-life is practically not affected by the electronic  bridge mechanism \cite{EB1}. It is in a reasonable agreement with the values obtained in the solid state experiments \cite{Th-wn0,Th-wn,Th-wn1,Hiraki,Zhang,Kraemer,Tkalya,Morgan,Perera}  after introducing the solid state corrections.  We can use this lifetime and factor $(1+\tilde \beta) \approx 1.7$  to conclude that half-life in Th III is  $T_{1/2}=820^{+350}_{-180}$ s .
%\section{Quantum information and search for new physics.}

%\begin{table} 
%  \caption{\label{t:hfs-E1} Amplitudes and rates of spontaneous decay for
%the electric dipole transition induced by the hyperfine interaction between
%the excited state at $E$=63 cm$^{-1}$, $J=2$ and the ground state.
%Nucleus is assumed to be in its ground state.} 
%\begin{ruledtabular}
%\begin{tabular}{cccc}
%$F_a$ &$F_b$ &$A_{ab}$ & $R_{ab}$ \\
%   &       & [a. u.]    &[s$^{-1}$] \\
%\hline
% 2.5 &  1.5 &    8.7[-6] &  9.8[-12] \\ 
% 2.5 &  2.5 &    1.1[-5] &  1.1[-11] \\ 
% 2.5 &  3.5 &   -6.7[-6] &  2.8[-12] \\ 
% 3.5 &  2.5 &   -1.1[-5] &  9.7[-12] \\ 
% 3.5 &  3.5 &   -2.2[-5] &  3.2[-11] \\ 
% 4.5 &  3.5 &    5.5[-7] &  2.0[-14] \\ 
%\end{tabular}
%\end{ruledtabular}
%\end{table}

\begin{table*} 
  \caption{\label{t:q} Parameters of three low-lying states of Th~III relevant to the  search for new physics. $\alpha_0$ is the  static dipole polarizability needed for estimation of the BBR shift, $Q$ is the  quadrupole moment of the state. 
  Enhancement of the variation of two atomic frequencies of the transitions between different configurations due to variation of the fine structure constant.} 
\begin{ruledtabular}
\begin{tabular}{lll rr cccc rr}
&\multicolumn{1}{c}{Conf.} &
\multicolumn{1}{c}{$J^p$} &
\multicolumn{1}{c}{$E_{\rm NIST}$} &
\multicolumn{1}{c}{$E_{\rm calc.}$} &

\multicolumn{1}{c}{$\alpha_{0}$} &
\multicolumn{1}{c}{$Q$} &
\multicolumn{1}{c}{$\langle a||T_0^{(2)}||a\rangle$} &
\multicolumn{1}{c}{$\langle a|H_K|a\rangle$} &

\multicolumn{1}{c}{$q$} &
\multicolumn{1}{c}{$K$} \\
&&&\multicolumn{1}{c}{[cm$^{-1}$}]& 
\multicolumn{1}{c}{[cm$^{-1}$}]& 

\multicolumn{1}{c}{[a.u.]}& 
\multicolumn{1}{c}{[a.u.]}& 
\multicolumn{1}{c}{[a.u.]}& 
\multicolumn{1}{c}{[a.u.]}& 

\multicolumn{1}{c}{[cm$^{-1}$}]& \\
\hline
GS  & $5f6d$ & $ 4^-$ &    0    &    0     & 13 & -4.5 & 52 &  1.0 & 0       &     0 \\
W1  & $6d^2$ & $2^+$ &  63    & 3056 & 38 & 0.22 & 3.0 & 1.5  & -33500 & -1060 \\
W2  & $5f7s$ & $3^-$  &  2527 & 2703 &      &         &       &        & -24160 &  -19 \\

\end{tabular}			
\end{ruledtabular}
\end{table*}

\section{Applications of Th III ion not related to nuclear clock} 
\subsection{Properties and possible application of metastable electronic state in Th III ion}
 As it was mentioned above,  Th III ion has  a very low lying (63~cm$^{-1}$) metastable state which has extremely large lifetime and can be considered as a second ground state. This doublet of states can be used for quantum information processing. In $^{229}$Th isotope the leading channel of decay is the E1 transition to the ground state is mediated by the hyperfine interaction. 
 The amplitude is given by
\begin{equation}\label{e:B}
A_{ab} = \sum_n\frac{\langle b|T_k|n\rangle\langle n|E1|a\rangle}{\epsilon_b-\epsilon_n}  
+ \sum_n\frac{\langle b|E1|n\rangle\langle n|T_k|a\rangle}{\epsilon_a-\epsilon_n} 
\end{equation}
%\begin{eqnarray}\label{e:B}
%A_{ab} &=& \left(\sum_n\frac{\langle b|T_k|n\rangle\langle n|E1|a\rangle}{\epsilon_b-\epsilon_n}  \right. \\
%&+&\left. \sum_n\frac{\langle b|E1|n\rangle\langle n|T_k|a\rangle}{\epsilon_a-\epsilon_n} \right) \nonumber
%\end{eqnarray}
where $T_k$ is the operator of the hyperfine interaction (as in (\ref{e:g2a})). Calculations show that corresponding rate of the spontaneous decay is $\sim 10^{-11} \ {\rm s}^{-1}$ %(see Table \ref{t:hfs-E1} ). 
The decay rate is small due to the small value of the transition frequency (the rate $\sim \omega^3$). Note however that this  $\omega^3$ factor is canceled by $\omega^3$ in the denominator  of the excitation rate, formula (\ref{e:ab-ba}).

It might be advantageous to use  the natural $^{232}$Th isotope instead of  $^{229}$Th. It has zero nuclear spin, eliminating the E1 amplitude induced by the  hyperfine interaction. The electronic  metastable state is connected to the ground state via an extremely  weak M2 transition with a lifetime $\sim 10^{10}$ years. This transition can be open by applying an external magnetic field. 
In this case, the transition amplitude is given by (\ref{e:B}), with $T_k$ replaced by the $M1$ operator and the amplitude multiplied by the  magnetic field strength $B$. 
Using calculated E1 and M1 
matrix elements for even and odd states with $J=3$ and experimental energies, we get 
he rate of  spontaneous decay 
$T_{ab} \approx 2 \times 10^{-5}~\rm{s}^{-1} \rm{T}^{-2}$.
 The rate of induced excitation can be estimated  using Eq. (\ref{e:ab-ba}).
Transition between the doublet of the ground states may also be organised via  E1  excitation and subsequent decay of higher states.  

\subsection{Violation of the Local Lorentz Invariance and Einstein Equivalence Principle} 

Ground state of the Th~III ion can be used in search for the  local Lorentz invariance (LLI) violation,  while the M2 transition can be used in search for the Einstein's equivalence  principle (EEP) violation. Corresponding Hamiltonian can be written as \cite{Kostelecky1,Kostelecky2,Kostelecky3,Hohensee,Nature}
\begin{equation}\label{e:LLI}
\delta H = -\left( C_0^{(0)} - \frac{2U}{3c^2}c_{00}\right)\frac{\mathbf{p}^2}{2} -\frac{1}{6}C_0^{(2)}T_0^{(2)},
\end{equation}
where $\mathbf{p}$ is the electron momentum operator, $c$ is the speed of light, $U$ is the gravitation potential, $C_0^{(0)}$, $c_{00}$ and $C_0^{(2)}$ are unknown constants to be found from measurements.
First term in (\ref{e:LLI}) violates the EEP via dependence of atomic frequencies on  time of the year caused by varying the Sun's gravitational potential. The change is periodical with minimum or maximum in January and July. To link the change to the unknown constant $c_{00}$, one should perform the calculations of the matrix elements of the $\mathbf{p}^2$ operator. The calculations must be relativistic since in the  non-relativistic limit all atomic frequencies change at the same rate and the effect is unobservable~\cite{EEP}. The relativistic form of the operator of kinetic energy is $H_K=c\gamma_0\gamma^jp_j$. It is convenient to present the result in terms of the relativistic factor $R$, which describes the deviation of the energy shift caused by the kinetic energy operator from the value given by the virial theorem~\cite{EEP}
\begin{equation}\label{e:R}
R= - \frac{\Delta E_a - \Delta E_b}{E_a-E_b}.
\end{equation}
Then $\Delta \omega/\omega = R\frac{2}{3}c_{00}\Delta U/c^2$. The calculated values of $\Delta E_a=\langle a|H_K|a\rangle$ and $\Delta E_b=\langle b|H_K|b\rangle$ for the ground and first excited states of Th~III are presented in Table~\ref{t:q}.  We  find an exceptionally  large relativistic factor, $R \approx -1700$,  due to the small energy denominator in  Eq. (\ref{e:R}). 
%Thus, the sensitivity of the frequency of the M2 transition to the change of the gravitation potential is strongly enhanced.
 Note that the values of the relativistic factors $R$ in systems studied previously were not high, $R \sim 1$~\cite{EEP,Yb}. 
This suggests that the Th~III ion may  be the most promising  system  for studying  the EEP violation.

Second term in (\ref{e:LLI}) causes LLI violation via dependence of the energy intervals between states with  different projections $M$ of the total atomic angular momentum $J$ on the system orientation, e.g. due to the Earth rotation. The non-relativistic form of the tensor operator $T_0^{(2)}$ is $\mathbf{p}^2 - 3p_z^2$, while the relativistic operator is $c\gamma_0(\gamma^jp_j-3\gamma^3p_3)$.
It was formulated in Ref.~\cite{Nature} that there are at least two conditions for the effect to be large: (a) long living state, (b) large matrix element which can be found in states with open $4f$ or $5f$ shells. Both of these conditions are satisfied for the ground state of Th~III.
Moreover, using the ground state is an advantage compared, e.g., to the Yb$^+$ ion, where  the effect is zero for the ground state.
Table~\ref{t:q} presents the values of the reduced matrix elements of the tensor operator $T_0^{(2)}$ for the ground and first exited state of Th~III. Note that the value for the ground state is only 2 to 3 times smaller than the value of the matrix elements for excited states of Yb$^+$, which have holes in the $4f$ subshell~\cite{Nature}.   

\subsection{Effect of variation of the fine structure constant}

Finally, the M2 transition can be used to search for the time variation of the fine structure constant $\alpha$. 
The calculated values of the relativistic energy shift $q$ and enhancement factor $K$ (see appendix) for two transitions between the ground state (GS) and two excited states,  labeled  W1 and W2, are presented in Table~\ref{t:q}.  
Note that all values of $q$ and $K$ are large and negative, in  agreement with an analytical estimate  (\ref{e:Delta}).
The value of $K$ for the M2 transition ($|K| \sim 10^3$) is several orders of magnitude larger than  in other   atomic clock systems  (see, e.g. \cite{Webb,WebbPRA,Safronova,CJP1,CJP2}). The only  exemption is Dy atom, which has a transition with an extremely large  enhancement factor ($K \sim 10^6 - 10^8$)~\cite{Dy-t}. However, this Dy transition is  not a clock transition,  and the sensitivity of measurements is  fundamentally limited by the transition linewidth ~\cite{Dy-E}. 
  
The sensitivity of the M2 transition in Th~III to the variation of $\alpha$  can be compared to  the recently estimated sensitivity of the nuclear transition
% frequency to the variation of $\alpha$, '
$K=5900(2300)$ \cite{nuc-alpha-dot,nuc-alpha}. 
%Note that 
If the nuclear  frequency is measured against the atomic frequency, the total sensitivity is further enhanced
\begin{equation}\label{e:K}
\frac{\delta \omega_N}{\omega_N} - \frac{\delta \omega_{M2}}{\omega_{M2}} \approx 7000 \frac{\delta \alpha}{\alpha}.
\end{equation}
For the further estimations, we have calculated static dipole polarizabilities and quadrupole moment for the both atomic clock states. The results are presented in Table~\ref{t:q}. Using the values of the polarizabilities, we estimate the black body radiation shift being smaller than 1~Hz. 
%The relative shift $\sim 10^{-13}$. 
The quadrupole moments for  the both clock states of Th~III are smaller than those in the clock states of Yb~\cite{Yb}. This reduces sensitivity to stray electric  fields.  Note that the corresponding frequency shift can be suppressed by averaging over projections of the total angular momentum (see e.g. Ref.~\cite{Yb}). 
%\acknowledgments

\section{Conclusions}

%We study the effect of the electronic bridge process on nuclear excitation in the $^{229}$Th~III and found that the nuclear excitation rate can be strongly enhanced, up to 300 times,  mediated by certain electronic transitions. The enhancement is achieved by proper choices of laser frequencies in a two-step process of atomic excitation, accompanied by energy transfer from electrons  to  nuclear excitation.

We have investigated the effect of the electronic bridge (EB) process on nuclear excitation in  $^{229}$Th~III  ion  and found that the nuclear excitation rate can be substantially enhanced - by up to $10^4$ times -  via  specific  electronic transitions. This enhancement is achieved through a two-step atomic excitation process,  followed by the transfer of most of the electron shell  energy to the nucleus, achieved by tuning the laser frequencies to excite both the nucleus and the final electronic state simultaneously.

 We also considered the  EB enhancement in a one-step excitation of  $^{229}$Th~III  ion  by a single photon.

The EB process also reduces  the nuclear isomeric state lifetime by about 1.7 times.

Our analysis of the low-lying electronic metastable state in Th~III  ion, which is connected to the ground state via a weak M2 transition, indicates its potential utility for quantum information processing. We also calculated possible signals of local Lorentz invariance and Einstein equivalence principle violations, and effect of  temporal variation of the fine-structure constant $\alpha$, which could result from interactions with dark matter or dark energy fields. These effects are substantially enhanced in Th~III ion, and combining measurements of both nuclear and atomic frequencies could further amplify their detectability.

In addition, this metastable electronic state may facilitate searches for atomic transition induced by axion dark matter field, as well as measurement of the weak nuclear quadrupole moment. We plan to present calculations of corresponding effects in future publications.

This work was supported by the Australian Research Council Grant No. DP230101058. We are grateful to Ekkehard Peik and Feodor Karpeshin  for a useful discussions.

\appendix 
\section{Method of calculation}

We use the relativistic Hartree-Fock (RHF) method and the combination of the configuration interaction with the single-double coupled cluster (CI+SD) method~\cite{CI+SD} to calculate two-electron valence states of the Th~III ion. To calculate transition amplitudes, we use the time-dependent Hartree-Fock method~\cite{TDHF}, which is equivalent to the well-known random-phase approximation (RPA).

The calculations start from the closed-shell Th~V ion. The single-electron basis states for valence electrons are calculated in the field of the frozen core using the B-spline technique~\cite{B-spline}. The SD equations are first solved for the core, then for the valence states~\cite{CI+SD}.
This leads to creation of the one-electron  and two-electron correlation operators $\Sigma_1$ and $\Sigma_2$, which are used in the CI calculations.
Solving the RPA equations for valence states leads to the effective operators of an external field, which are used to calculate matrix elements between valence states. The accuracy of the calculations is illustrated further in the text by comparing calculated energies and $g-$factors with experiment (see Table~\ref{t:ener}).

The sensitivity of the frequency of the M2 transition to the variation of $\alpha$ is strongly enhanced due to high $Z$ and  due to the fact that the transition is between states of different configurations. 
The latter can be explained in the following way. The relativistic energy shift of a single-electron state is given by~\cite{Webb,WebbPRA}
\begin{equation}\label{e:Delta}
\Delta_n \approx \frac{E_n(Z\alpha)^2}{\nu}\left[\frac{1}{j+1/2}-0.6\right].
\end{equation}
Here $\nu$ is the effective principal quantum number, $\nu=1/\sqrt{-2E_n}$, $j$ is the total angular momentum of electron orbital.
One can see from (\ref{e:Delta}) that the maximum frequency shift due to  $\alpha$ varition ($\delta \omega \approx \Delta_{n_1} - \Delta_{n_2}$) can be achieved for transitions with the largest $\Delta j$. 

The Th~III is a unique atomic system in which the single-electron energies of the $5f$, $6d$ and $7s$ states are very close. 
Therefore, transition between low-lying states of Th~III are usually either $5f-6d$ or $6d-7s$ transitions.
%It was found in Refs.~\cite{Webb,WebbPRA} that large $\Delta j$ in single-electron transitions lead to large relativistic shifts.

To calculate the sensitivity of atomic frequencies to the variation of the fine structure constant we present them in a form
\begin{equation}\label{e:w0}
\omega(x) = \omega_0 + q\left[x-1\right], \ \ x \equiv \left(\frac{\alpha}{\alpha_0}\right)^2,
\end{equation}
where $\omega_0$ and $\alpha_0$ are physical values of the frequency and fine structure constant respectively, $q$ is sensitivity coefficient to be found  by varying the value of $\alpha$ in computer codes and calculating the numerical derivative.
\begin{equation}\label{e:q}
q = \frac{\omega(+\lambda)-\omega(-\lambda)}{2\lambda},
\end{equation}
where $\lambda$ is a small parameter, we usually use $\lambda = 0.01$.
Parameter $q$ links variation of atomic frequency to the variation of $\alpha$
\begin{equation}\label{e:K}
\frac{\delta \omega}{\omega} = K \frac{\delta \alpha}{\alpha}.
\end{equation}
The dimensionless factor $K=2q/\omega$ is called {\em enhancement factor}.

%Matrix elements of the electric dipole operator, which are used in the calculations, are presented in Table~\ref{t:E1}.

%\section{Method of calculations}

\end{document}